%Paper: astro-ph/9306004
%From: MOORE@BKYAST.BERKELEY.EDU
%Date: Thu, 10 Jun 1993 8:43:45 -0700 (PDT)

% Ok, here is the letter. It was accepted for publication in Ap.J.Letters
% on June 1st 1993. Further comments welcomed.
% One Figure is not included. Send me an email at moore@bkyast.berkeley.edu
% and I will send a copy to you.
% One Table is included at the end of this file *** this is LATEX format ***
% Cut this out and print it seperately. (lines 767 to end).
% If none of this works, send me an email and I will send you a preprint.
% The large marcro at the beginning of this file formats the paper into a copy
% of Ap.J.Letters. You might have to search for voffset and hoffset and change
% these to your local printer settings. Good luck!
\def\etal{\it et al. \rm}
\def\kms{km s$^{-1}$}

% define a greater than or approx. equal to sign
\def\gsim{ \lower .75ex \hbox{$\sim$} \llap{\raise .27ex \hbox{$>$}} }
\def\lsim{ \lower .75ex \hbox{$\sim$} \llap{\raise .27ex \hbox{$<$}} }
\def\pp{\noindent\parshape 2 0truecm 16truecm 1truecm 15truecm}
%
%  \pp is a simple definition to define a paragraph shape in
%      which the first line is not indented, but subsequent lines are.
%      suitable for references and figure captions.
% useful for references
\def\pp{\noindent\parshape 2 0truecm 16.5truecm 2truecm 14truecm}

% produces <~ or >~ signs
\def\spose#1{\hbox to 0pt{#1\hss}}
\def\simlt{\mathrel{\spose{\lower 3pt\hbox{$\mathchar"218$}}
     \raise 2.0pt\hbox{$\mathchar"13C$}}}
\def\simgt{\mathrel{\spose{\lower 3pt\hbox{$\mathchar"218$}}
     \raise 2.0pt\hbox{$\mathchar"13E$}}}
%
% define a greater than or approx. equal to sign
%
\def\gsim{ \lower .75ex \hbox{$\sim$} \llap{\raise .27ex \hbox{$>$}} }
\def\lsim{ \lower .75ex \hbox{$\sim$} \llap{\raise .27ex \hbox{$<$}} }
%
%******
% THIS FILE CONTAINS MACROS TO MIMIC THE LAYOUT OF ApJ. LETTERS.
% THE FONT ETC. IS NOT IDENTICAL, BUT IT HAS BEEN TESTED ON A
% REAL ApJ. TEXT TO GIVE (AS NEAR AS POSSIBLE) THE SAME LENGTH.
% FOLLOW THE FORMAT AT THE END OF THIS FILE.

%(Written by Dan Plonsey, with improvements by Douglas Scott)

%(More improvements to be made - ApJ. uses a smaller inter-sentence
%space than the TeX standard - the spaces left between sections seem
%to vary a lot - tables and figures could be added!)

%First redefine the fonts to be am*** instead of cm***, so that the fonts
% of the required magnification (0.96, ie ***.288pk) are available
\catcode`@=11
 \font\tenrm=amr10 % roman text
  \font\sevenrm=amr7
  \font\fiverm=amr5
 \font\teni=ammi10 % math italic
  \font\seveni=ammi7
  \font\fivei=ammi5
 \font\tensy=amsy10 % math symbols
  \font\sevensy=amsy7
  \font\fivesy=amsy5
 \font\tenex=amex10 % math extension
 \font\tenbf=ambx10 % boldface extended
  \font\sevenbf=ambx7
  \font\fivebf=ambx5
  % typewriter
 \font\tensl=amsl10 % slanted roman
 \font\tenit=amti10 % text italic
 \skewchar\teni='177 \skewchar\seveni='177 \skewchar\fivei='177
 \skewchar\tensy='60 \skewchar\sevensy='60 \skewchar\fivesy='60
 \textfont0=\tenrm \scriptfont0=\sevenrm \scriptscriptfont0=\fiverm
  \def\rm{\fam\z@\tenrm}
 \textfont1=\teni \scriptfont1=\seveni \scriptscriptfont1=\fivei
  \def\mit{\fam\@ne} \def\oldstyle{\fam\@ne\teni}
 \textfont2=\tensy \scriptfont2=\sevensy \scriptscriptfont2=\fivesy
  \def\cal{\fam\tw@}
 \textfont3=\tenex \scriptfont3=\tenex \scriptscriptfont3=\tenex
 \newfam\itfam \def\it{\fam\itfam\tenit} % \it is family 4
  \textfont\itfam=\tenit
 \newfam\slfam  % \sl is family 5
  \textfont\slfam=\tensl
 \newfam\bffam \def\bf{\fam\bffam\tenbf} % \bf is family 6
  \textfont\bffam=\tenbf \scriptfont\bffam=\sevenbf
  \scriptscriptfont\bffam=\fivebf
 \newfam\ttfam  % \tt is family 7
 \normalbaselineskip=12pt
 \setbox\strutbox=\hbox{\vrule height8.5pt depth3.5pt width 0pt}
 \normalbaselineskip\rm
\catcode`@=12

%Slightly decrease the size of the fonts
\magnification=960

%Load the special fonts
\font\smc=amcsc10 at 8.77truept
\font\tinyrm=amr8
\font\verytinyrm=amr9 at 6truept
\font\tinybf=ambx8
\font\tinyit=amti8
\font\headsmc=amcsc10 at 6.67truept
\font\headbf=ambx10 at 7.3truept
\font\headrm=amr10 at 7.3truept

%Now decrease the normal spaces left in maths formulas to mimic ApJ
\thinmuskip=2mu
\medmuskip=3mu plus 1mu minus 3mu
\thickmuskip=4mu plus 4mu

%Set up a tiny copyright symbol
\def\fixcircle{\sevensy\char"0D}
\def\tcopyright{{\ooalign{\hfil\lower.05ex\hbox{C}\hfil\cr\cr\fixcircle}}}

%
%  macro for invoking today's date (when TeX is run on your file)
%
\def\today{\number\year\space \ifcase\month\or 	January\or February\or
	March\or April\or May\or June\or July\or August\or September\or
	October\or November\or December\fi\space \number\day}

%Now make the headers and footers, making sure they aren't in doublecols.
\newdimen\fullhsize
\fullhsize=7.53truein
\def\fullline{\hbox to\fullhsize}
\def\makeheadline{\vbox to 0pt{\vskip-22.5pt
 \fullline{\vbox to8.5pt{}\the\headline}\vss}\nointerlineskip}
\def\makefootline{\baselineskip=12pt\fullline{\the\footline}}
\def\rightheadline{\rm No.~1,\ \number\year \hfil \shorttitle \hfil L\folio}
\def\leftheadline{\rm L\folio \hfil \shortauthor \hfil Vol.\ 999}
\def\asthead{\vbox{\noindent{\headsmc The Astrosynthetic Journal,
 \headbf 999\headrm :L1--L4, \number\year\ February 31\hfil}\vskip3pt
 \noindent{\verytinyrm\tcopyright 1985, Berkeley
 Astronomy Department. All rights reserved. Printed in
U.S.A.\hfil}}}

\headline={\ifnum\pageno=1 \asthead \else{\ifodd\pageno\rightheadline
 \else\leftheadline\fi}\fi}
\footline={\ifnum\pageno=1 {\hfil L1\hfil} \else{\hfil}\fi}

%Set up all the sizes
\voffset=.95truein
\vsize=9.3truein
\hoffset=.45truein
\hsize=7.53truein %makes lines too wide, but approx. same no. chars per line
\parindent=.12truein
\parskip=0pt
\baselineskip=11truept
\baselineskip=10.5truept

%Define the section macros
\def\Instaddress{Astronomy Department. University of California, Berkeley,
CA 94720}
\def\beginabstract{\centerline{\tinyrm\Instaddress}

\centerline{\tinyrm Center for Particle Astrophysics. University of
California, Berkeley.}

 \vskip-2pt\centerline{\tinyit Written \today}
 \begingroup\leftskip=.5truein\rightskip=.5truein\parskip=3truept
 \medskip\smallskip\centerline{ABSTRACT}}

\def\subjectheadings#1{\vskip-2pt\par\noindent{\it Subject headings:\/} #1}
\def\endabstract{\medskip\smallskip\endgroup\begindoublecolumns}
\def\beginref{\enddoublecolumns\medskip\smallskip\centerline
  {\smc references}\begindoublecolumns\begingroup\baselineskip=7.9pt
  \parindent=0in\let\bf=\tinybf\let\it=\tinyit\let\rm=\tinyrm\rm}
\def\endref{\smallskip\endgroup\vfil\enddoublecolumns\parskip=12truept
  \parindent=0in}
\def\section#1{\medskip\penalty -500\centerline{\smc #1}\medskip}
%
% Some useful symbols
%

\def\>{$>$}
\def\<{$<$}

\def\simlt{\lower.5ex\hbox{$\; \buildrel < \over \sim \;$}}
\def\simgt{\lower.5ex\hbox{$\; \buildrel > \over \sim \;$}}
\def\sqr#1#2{{\vcenter{\hrule height.#2pt
      \hbox{\vrule width.#2pt height#1pt \kern#1pt
         \vrule width.#2pt}
      \hrule height.#2pt}}}

%  These provide a simple means of producing references in ApJ style.
%
\def\pp{\par\hangindent=.125truein \hangafter=1}
\def\apjref#1;#2;#3;#4{\pp #1, {\it #2}, {\bf #3}, #4.}
\def\ref#1;#2;#3;#4{\pp #1, {\it #2}, {\bf #3}, #4}
\def\book#1;#2;#3{\pp #1, {\it #2}, #3}
\def\rep#1;#2;#3{\pp #1, #2, #3}

	%underline some text
%
%*****
% THIS FILE IS A COLLECTION OF MACROS TO BE USED FOR CREATING DOUBLE-COLUMN
% OUTPUT. TO USE, TYPE \begindoublecolumns AT THE BEGINNING OF THE SECTION,
% THEN TYPE \enddoublecolumns AT THE END.
% THESE MACROS ARE TAKEN FROM tugboat vol.6, no.1., p.30. THEY ARE A REVISED
% VERSION OF THE MACROS IN the texbook, p.417
%
% Modified 6/19/91 DJP -- semi-fixed the page number problem.
%
\tolerance=10000
\newdimen\colwidth \newdimen\bigcolheight
\newdimen\pagewidth \newdimen\pageheight
%
% THE \colwidth PARAMETER CONTROLS COLUMN WIDTH.
% THE \bigcolheight PARAMETER SHOULD BE SET TO  2 * \vsize
%
\colwidth=\hsize
  \advance\colwidth by -.35truein
  \divide\colwidth by 2
\bigcolheight=\vsize
  \advance\bigcolheight by \vsize
\newdimen\savevsizea \savevsizea=\vsize \advance\savevsizea by 24pt
\newdimen\savevsize \savevsize=\vsize
\newdimen\savehsize \savehsize=\hsize
\def\makefootline{\baselineskip=24pt\hbox to \savehsize{\the\footline}}
\pagewidth=\hsize \pageheight=\vsize
\def\onepageout#1{\shipout\vbox{
    \offinterlineskip
    \makeheadline
    \vbox to\savevsizea{#1\vfill}
    \makefootline}
    \advancepageno}
\output{\onepageout{\unvbox255}}
\newbox\partialpage
\def\begindoublecolumns{\begingroup
  \output={\global\setbox\partialpage=\vbox{\unvbox255}}\eject
  \output={\doublecolumnout} \hsize=\colwidth \vsize=\bigcolheight
  \advance\vsize by -2\ht\partialpage}
\def\enddoublecolumns{\output={\balancecolumns}\eject
  \global\output={\onepageout{\unvbox255}}
  \global\vsize=\savevsize
  \endgroup \pagegoal=\vsize}
\def\doublecolumnout{\dimen0=\pageheight
  \advance\dimen0 by-\ht\partialpage \splittopskip=\topskip
  \setbox0=\vsplit255 to\dimen0
  \setbox2=\vsplit255 to\dimen0
  \onepageout\pagesofar
  \global\vsize=\bigcolheight
  \unvbox255 \penalty\outputpenalty}
\def\pagesofar{\unvbox\partialpage
  \wd0=\hsize \wd2=\hsize \hbox to\pagewidth{\box0\hfil\box2}}
\def\balancecolumns{\setbox0=\vbox{\unvbox255} \dimen0=\ht0
  \advance\dimen0 by\topskip \advance\dimen0 by-\baselineskip
  \divide\dimen0 by2 \splittopskip=\topskip
  {\vbadness=10000 \loop \global\setbox3=\copy0
    \global\setbox1=\vsplit3 to\dimen0
    \ifdim\ht3>\dimen0 \global\advance\dimen0 by1truept \repeat}
  \setbox0=\vbox to\dimen0{\unvbox1}
  \setbox2=\vbox to\dimen0{\unvbox3}
  \global\output={\balancingerror}
  \pagesofar}
\newhelp\balerrhelp{Please change the page
                        into one that works.}
\def\balancingerror{\errhelp=\balerrhelp
        \errmessage{Page can't be balanced}
        \onepageout{\unvbox255}}
%*****
%\input twocolumns.tex	%DEFINES TWO-COLUMN FORMAT AND OUTPUT ROUTINE
%\Noline
%******
%	\input apjlet
%Define the short form of title and author for the page headers
	\def\shorttitle{BLACK HOLES AND GLOBULAR CLUSTER DISRUPTION}
	\def\shortauthor{BEN MOORE}
	\null
	\vskip .45truein
%If the institution's address is other than Astronomy Department. University
% of California, Berkeley, CA 94720, then define \Instaddress:
%       \def\Instaddress{Institution, address}
%Put in the title and author

	\centerline{AN UPPER LIMIT TO THE MASS OF BLACK HOLES}
	\centerline{IN THE HALO OF OUR GALAXY}
	\medskip
	\centerline{\smc {\bf Ben Moore}}
	\beginabstract

If massive black holes constitute the dark matter in the halo surrounding the
Milky Way, the existence of low mass globular clusters in the halo suggests an
upper limit to their mass, $M_{_{BH}}$. We use a combination of the impulse
approximation and numerical simulations to constrain $M_{_{BH}} \lsim
10^3M_\odot$, otherwise several of the halo globular clusters would be heated
to
disruption within one half of their lifetime. Taken at face value, this
constraint is three orders of magnitude stronger than the previous limit
provided by disk heating arguments. However, since the initial mass function of
clusters is unknown, we argue that the real constraint is at most, an order
of magnitude weaker. Our results rule out cosmological scenarios, such as
versions of the Primordial Baryonic Isocurvature fluctuation model, which
invoke
the low Jeans mass at early epochs to create a large population of black
holes of mass $\sim 10^6M_\odot$.

\subjectheadings{Cosmology: dark matter \ \------ \ Globular clusters:
general \ \------ \ Galaxy: halo}
	\endabstract		%THIS WILL BEGIN DOUBLECOLUMN FORMAT
%For section headings, do this (in lower case)
\medskip

	\section{1. INTRODUCTION}

The Local Group of galaxies provides a unique laboratory for probing the nature
and distribution of dark matter on scales of 100's of kpc. Dynamical analyses
of
the local mass distribution, using the radial velocities and `true' distances
of
the Milky Way's satellites and Andromeda, suggest that the density of material
in our halo decreases as slowly as the inverse square of the distance out to at
least 70kpc (for a review, see Fich and Tremaine 1992). This implies a
significant amount of dark matter at large galactocentric distances. The key
question, still unanswered after several decades of investigation, is: What
constitutes the dark matter in the halos of galaxies? Many baryonic and
non-baryonic candidates for the dark matter have been proposed. The three main
contenders are: black hole remnants of very massive stars, low mass stars such
as white dwarfs or `Jupiters', or some exotic species of particle surviving
from
the big bang (for a review, see Rees 1987).

Theoretical interest in massive black holes stems from the fact that the Jeans
mass at high redshift is $1.3\times 10^6 M_\odot \Omega_{_B} \Omega_o {}^{-1/2}
h^{-1}$ (Dicke and Peebles 1968). Here $\Omega_{_B}$ and $\Omega_o$ are the
baryonic and total density parameters and $h$ is the Hubble constant in units
of
100\kms\ Mpc$^{-1}$. Hence an early burst of massive star formation could
conceivably form a large population of black holes in the mass range $2\times
10^5M_\odot$ to $3\times 10^6M_\odot$, where the lower limit would
correspond to theoretical prejudice. Some versions of the Primordial Baryonic
Isocurvature fluctuation scenario for structure formation assume that a large
population of black holes of mass $\sim 10^6M_\odot$ form at high redshifts.
These objects are predicted to constitute the dynamically inferred dark matter
in the halos of galaxies (Peebles 1987, Cen \etal 1993).

Lacey and Ostriker (1985) proposed that black holes in our halo of mass $\sim
2\times 10^6M_\odot$ may be responsible for heating the disk of our Galaxy.
These authors also highlight the problems associated with objects of this mass
constituting the dark matter. In particular, dynamical friction would cause the
black holes within a few kpc to sink to the center of the Galaxy. In addition,
black holes would accrete matter as they move through the disk and violate
background light constraints (Carr 1979). A tight lower limit to $M_{_{BH}}$ is
given by Carr \etal (1984), who demonstrate that $M_{_{BH}}>100M_\odot$,
otherwise the interstellar medium would be significantly enriched with heavy
elements from exploding stars which do not collapse directly into black holes.
Several observational projects are underway which aim to constrain the
abundance
of massive compact halo objects (MACHOs) ({\it i.e.} Bennet 1992, Magneville
1992). These experiments are searching for lensing of distant stars by MACHOs
and can be sensitive in the range $10^{-6}M_\odot \sim 10^{4}M_\odot$.

In this {\it letter} we argue that the present upper limit to the mass of black
holes in our halo, MACHOs, or clusters of these objects (Carr and Lacey 1987),
can be reduced to preclude objects of the Jeans mass from constituting the dark
matter, perhaps allowing the entire window on baryonic dark matter candidates
to
be observed within the next few years. We proceed in the same spirit as Spitzer
(1958) who utilised the impulse approximation to calculate the disruption
timescale, $\tau_{_D}$, of star clusters in the disk by encounters with giant
molecular clouds. In a similar fashion, encounters with a halo population of
massive black holes would cause the internal energy of the halo globular
clusters (hereafter GCs) to increase with time, hence mass would be lost as
loosely bound stars escape past the tidal radius.

Wielen (1986) used Spitzer's formula to consider the heating effects of black
holes of mass $M_{_{BH}} = 3\times 10^6M_\odot$ and concluded that many GCs
could have been disrupted from this mechanism. In this paper we consider more
carefully several important effects which Wielen ignored, and apply our results
to a specific set of well observed GCs. In particular, we use numerical
simulations to calculate the energy input from penetrating encounters as well
as
to study the regime over which the impulsive approximation is valid. These
areas
have generally been ignored during applications of the impulse approximation.
We
also consider a distribution of encounter velocities and we use King models to
study the change in the GCs structural parameters. Our results are supported by
self consistent numerical simulations which we use use to study the disruption
process. We find that Wielen's neglect of these effects lead $\tau_{_D}$ to be
underestimated by two orders of magnitude.

\vfil\eject

	\section{2. THE IMPULSE APPROXIMATION}

\smallskip
\centerline{2.1. {\it Direct and distant encounters}}
\smallskip

A black hole moving with velocity $V_{_{BH}}$, will
cause a change in velocity of a star:
$$
\eqalignno{
\delta V &= 2GM_{_{BH}}/bV_{_{BH}}\ ,  &(1)\cr
}
$$
towards the point of closest approach of the perturber with
impact parameter $b$.
If $b^2 \gg \overline{r^2}$, the mean square radius, then the encounter
causes a net increase in the
kinetic energy, $\Delta E_{distant}$, of a stellar system of mass $M_{_{GC}}$,
which which can be calculated exactly (Spitzer 1958):
$$
\eqalignno{
\Delta E_{distant} &= {{4G^2M_{_{BH}}^2 M_{_{GC}}
\overline{r^2}}\over{V_{_{BH}}^2b^4}} \ . & (2)\cr
}
$$

If $b=0$ then the total change in kinetic energy, $\Delta E_{direct}$, can be
calculated by integrating the mean square velocity change over the system's
surface density, $\sum (R)$ ({\it i.e.} $\sim$ Binney and Tremaine 1987):
$$
\eqalignno{
\Delta E_{direct} &= \pi\int_0^\infty [\delta V(R)]^2 \sum (R) R dR\ . &(3)\cr
}
$$
The surface density of a GC can be well approximated by a King model specified
by its central density and concentration $c={\rm log}_{10}(r_t/r_c)$, where
$r_t$ is the radius where the surface density falls to zero and $r_c$ is the
core radius. Substituting for $\delta V$, we can write
the energy input for direct encounters:
$$
\eqalignno{
\Delta E_{direct} &= {{4\pi G^2 M_{_{BH}}^2}\over{V_{_{BH}}^2}}
\int_{b_{min}}^{b_{max}} {{\sum (R)}\over{R}} dR\ , &(4)\cr
}
$$
where $b_{min}$ and $b_{max}$ are the minimum and maximum impact parameters to
be considered. This integral diverges at small impact parameters hence we set
$b_{max}=r_t$ and define $b_{min}$ to be the minimum impact parameter which
would give stars an impulse velocity greater than the escape velocity, $v_{esc}
= \sqrt{2\Phi(0)}$, where $\Phi(0)$ is the central potential of the King model.
(Note that $b_{min}$ depends on both the mass and velocity of the black hole as
well as the structural parameters of the GC.)

We now have analytic estimates of the internal energy change for direct
and distant collisions. To interpolate between these extremes and to check
the validity of the impulse approximation, we use N-body simulations.

\medskip
\centerline{2.2 {\it Numerical simulations of penetrating encounters}}
\smallskip

We construct a GC with a projected density profile given by King's (1966)
prescription with isotropic orbits and core radius $r_c=4$pc, tidal radius
$r_t=30$pc and mass $M_{_{GC}}=10^3M_\odot$. For this model, $v_{esc}\sim
1.2$\kms\ and the central velocity dispersion, $\sigma \sim 0.4$\kms. We evolve
the GC in isolation with the TREE code (Barnes and Hut 1987) for several
crossing times in order to settle the transient stellar motions. The softening
parameter is 0.1pc and the timestep is set to $10^4$yrs which conserves energy
to better than 0.1\% over a Gyr.

First we calculate $\Delta E$ as a function of impact parameter:
\ \ We send black holes of mass $10^4 M_\odot$ and velocity 200\kms\ through
the
GC over a range of impact parameters. After several crossing times the GC has
time to virialise and distribute the additional kinetic energy into potential
energy. Stars which escape past the
tidal radius are removed from the simulation and we calculate the total energy
change after re-normalising to the center of velocity.

Figure 1 shows the results of this calculation. We plot the fractional change
in
kinetic energy against the impact parameter. We found that equation (2) is
accurate to within 10\% for impacts as close as $r_t$. When
$b=r_c$ the energy input is about 1\% of the binding energy,
$E_{bind}=-kGM_{_{GC}}^2/r_t$. In this equation, $k$ is a constant which
depends
upon the concentration of the GC; if $c=1$, then $k=1.09$. An excellent
empirical fit to this dependance is given by
$$
\eqalignno{
{{\Delta E(b)}} &= {{\Delta E(0)}\over{(1+(b/r_c))^4}}\ , &(5)\cr
}
$$
where $\Delta E(0)$ is the fractional energy change at $b=0$.
When $b=0$, $\Delta E(0) = 0.15 E_{bind}$, which is a few percent
higher than predicted using equation (4) with $b_{min}$ set to the escape
velocity. The reason for this discrepancy is that stars which receive impulse
velocities greater than $v_{esc}$, transfer additional energy
during star-star collisions as they leave the cluster (Chandrasekhar 1942).
{}From this equation it is clear that penetrating encounters, $b<r_t$, will
input
a great deal more energy than distant encounters.

The basic assumption in the impulse approximation is that the effective
encounter time is much smaller than the typical crossing time of the stellar
system. When the encounter is slow, the system will have time to deform
adiabatically and very little energy is transferred.
For a range of impact parameters and velocities, we send black
holes of fixed mass past the GC. We find that the energy input scales as
equations (4) and (5) to better than 5\%, provided that the crossing time of
the GC, $r_c/(2\sigma)$, is at least 10 times as long as that of the
black hole, $b/V_{_{BH}}$.

\medskip
\section{3. DISRUPTION TIMESCALES}
\smallskip

The subsequent evolution of a stellar system under the influence of
gravitational perturbations, mass loss and a tidal field is complex ({\it c.f.}
$\sim$ Chernoff and Shapiro (1987), Lee and Ostriker 1987, Aguilar \etal 1988).
Chernoff \etal (1986) perform detailed numerical and analytical analyses of the
effect of tidal perturbations on the evolution of a GC. They conclude that a GC
could either expand or contract, depending on its initial concentration and the
ratio of energy gain by direct heating, to potential gain from mass loss. The
critical concentration was observed to be constant, $c\sim 0.9$, over the
range of perturber masses, $10^4M_\odot \sim 10^6 M_\odot$.

A detailed investigation of the behaviour of each of the halo GCs is beyond the
scope of this {\it letter}, however, later we will illustrate these effects
using numerical simulations of a model GC with the critical concentration. As a
first approximation, we can simply sum the $\Delta E's$ from each encounter and
equate the disruption timescale, $\tau_{_D}$, with the time over which the GCs
internal energy changes by an amount equal to its binding energy.

The rate at which GCs encounter black holes can be written as $\dot{C} = 2\pi
\overline{n} V_c f(v) dv b db$, where $f(v)$ is the distribution of impact
velocities and $\overline{n}$ is the number density of perturbers. In our halo,
most of the mass is dark beyond 10kpc and the mass density at distance $r$ can
be calculated from $\rho(r) = V_c^2/4\pi G r^2$, where the circular velocity,
$V_c=220$\kms. For example, Pal~5 is at a distance $\sim 16$kpc and has a tidal
radius, $r_t\sim 76$pc, hence we would expect about 1500 penetrating
encounters per Gyr with $10^4M_\odot$ black holes.

A good approximation for $f(v)$ is an isotropic Maxwellian distribution with
dispersion $\sqrt{(3/2)}V_c \sim 270$\kms. The circular velocity of the GC
effectively increases the dispersion of impact velocities and the number of
encounters, resulting in a similar rate of energy input as if the cluster were
stationary. Circular orbits yield similar rates of energy input, however,
radial orbits lead to longer $\tau_{_D}$ since there would be no slow
velocity tail in the distribution of impact velocities. However, radial orbits
are unstable (Palmer and Papaloizou 1987), a bar quickly forms and the orbits
are isotropised, moreover, cosmological N-body simulations create halos with
typically isotropic orbital distributions (Frenk \etal 1988).

Column 1 of Table 1 gives the dynamical properties of several halo GCs with
galactocentric distances given in column 2. These data were taken mostly from
the compilations of Webbink (1984) and Trager \etal (1993). The core and tidal
radii are listed in columns 3 and 4. These are usually estimated directly from
star counts using photographic plates ({\it e.g.} Sandage and Hartwick 1977) or
from CCD images ({\it e.g.} Djorgovski 1986). Most of the GC mass estimates in
column 5 were calculated from the total visual magnitude of the clusters,
assuming $M_{_{GC}}/L_v = 1.6M_\odot/L_{\odot ,v}$ (Illingworth and Illingworth
1976). Clusters with measured stellar velocity dispersions have mass to light
ratios in the range 1 to 2 (Pryor \etal 1986), therefore $\tau_{_D}$
could be up to 25\% longer since
$\tau_{_D}$ increases roughly $\propto M_{_{GC}}$.
For the mass of AM~4 we use the upper limit given by Inman and
Carney (1987), and the distance of Pal~15 is taken from Seitzer and Carney
(1990).

We calculate $\tau_{_D}$ using Monte-Carlo techniques. For each GC we calculate
its $\sum(R)$. For a particular $M_{_{BH}}$ we integrate equation (4) to
calculate $\Delta E(0)$. We then choose a random impact parameter within 500pc
of the cluster and a random velocity drawn from $f(v)$. The impulse energy at
this impact parameter is calculated using equation (5). This is repeated $10^6$
times and the total energy input is summed and normalised to the expected
collision rate, $\dot{C}$. We calculate the minimum mass black hole which would
input an amount of energy equal to the GCs binding energy within 7 Gyrs
(columns
6 and 7 of Table 1). We choose this timescale because it is roughly half of the
typical age of the halo GCs and half of a Hubble time ($h=0.75$). AM~4 is most
susceptible to disruption; $M_{_{BH}} \lsim 7\times 10^2 M_\odot$. However,
since the structural parameters of AM~4 are poorly determined, a stronger
constraint is provided by the well observed cluster Pal~5, $M_{_{BH}}\lsim
1.1\times 10^3M_\odot$.

\medskip
	\section{4. DISCUSSION}
\smallskip
\centerline {4.1. {\it Numerical tests}}
\smallskip

As discussed in Section 3, the actual $\tau_{_D}$ of a particular GC will be
determined by how its structure responds to the energy input. Six of the halo
clusters in Table 1 have concentration less than the critical value, hence
these
clusters are expected to expand and disperse quite rapidly (at least for
perturbers with $M_{_{BH}}\gsim 10^4M_\odot$). To illustrate this effect, and
to
test our analytical estimates of $\tau_{_D}$, we perform several numerical
simulations of a GC whose concentration lies at the critical value, $c=0.9$.

At random time intervals we calculate a random trajectory through a sphere of
radius 500pc around the model GC. We calculate the closest approach of each
star
to the black hole and give the star a corresponding impulse
velocity calculated from equation (1) and subject to the assumptions determined
in Section 3. Any star which crosses the tidal radius is subtracted from
the simulation and a new tidal radius is calculated.
%using
%$r_t = r_{_G}(M_{_{GC}}/3M_{_G})^{1/3}$. Here, the mass of the Galaxy
%$M_{_G}=V_c^2r_{_G}/G$ within the orbit of the GC at distance $r_{_G}$, and
%we are ignoring the effect of $dM_{_G}/dr_{_G}$.
We ran several simulations using a King model similar to that described above,
with $M_{_{BH}}=10^3, 10^4$ and $10^5M_\odot$. Each run assumed an isotropic
distribution of perturbers with velocities drawn from a Gaussian of width
270\kms\ and $r_{_G}=30$kpc. For these parameters we expect
$\tau_{_D}$ to be $\sim 12.5, 2.1$ and $0.3$ Gyrs, respectively.

In each simulation with $10^5M_\odot$ black holes, the cluster disrupted quite
violently and within a very short timescale, less than 0.1 Gyrs. The clusters
evolved quite stochastically since $\tau_{_D}$ depends strongly on
a few close encounters. With $10^4M_\odot$ black holes the GCs lost all of
their
mass within 1.7Gyrs, a fraction of their crossing time, and again in a
timescale
shorter than predicted using the impulse approximation. One of these models
disrupted within half a Gyr. The reason for this rapid disruption, is that the
balance of energy input to mass loss caused the cluster to expand, and the rate
of mass loss increased with time. During the disruption of these clusters, the
projected density profiles appeared distorted and quite dissimilar to those of
the GCs we observe in the halo of our Galaxy.

We ran one simulation with $10^3 M_\odot$ perturbers. In this case, the GC was
heated quite slowly. For the first 5Gyrs, mass was lost at a uniform rate,
however, the concentration of the cluster gradually increased. The energy
transfer became less efficient, and the rate of mass loss decreased and became
proportional to the inverse of the time. After 12Gyrs, 25\% of the mass
remained
and the core radius had decreased to $\sim 1$pc and the concentration had
increased from 0.9 to well above unity. After 17Gyrs, only 10\% of the mass
remained within a radius of 3pc.

\smallskip
\noindent {4.2. {\it The initial mass function of globular clusters}}
\smallskip

It could be argued that the GCs which we observe today are the remnants of a
once larger population, most of which have already disrupted, perhaps forming
the observed halo population of stars. Although we cannot firmly rule out the
possibility of MACHOs of mass $\sim 10^3 M_\odot$, there are several plausible
arguments against more massive halo objects than this:

{\noindent (i)} \ Chernoff \etal (1986) predicted that shocks from disk
crossing
and interactions with giant molecular clouds would increase the rate of core
collapse since these processes increase the concentration of typical GCs.
Indeed, Djorgovski and King (1986) found that all of the GCs which show the
characteristic cusps indicating core-collapse, lie at galactocentric distances
less than 10kpc. The fact that we observe no core collapsed GCs in the halo
indicates either that these clusters are not suffering perturbations from black
holes, or that the initial distribution function of GCs was contrived so that
most clusters were born with a sharp cut-off in concentration at $c \sim 0.9$.

{\noindent (ii)} \ GC profiles appear symmetric and
relatively circular ({\it i.e.} White and Shawl 1987). From our numerical
simulations of GCs with massive perturbers, we noticed that during disruption,
the projected density profiles appear quite irregular on the sky \---
unlike any of the clusters in our halo which do not appear to be violently
disrupting.

{\noindent (iii)} \ $\tau_{_D}$ scales approximately as
$M_{_{GC}}/M_{_{BH}}$, therefore if $M_{_{BH}} \gsim$ the Jeans Mass, Pal~5
would disrupt within one tenth of a Gyr. For us to be observing this cluster at
this stage in its evolution is highly unlikely, moreover, this would imply that
its initial mass was greater than the most massive GCs which we observe today.
Furthermore, if $M_{_{BH}}\sim 10^6M_\odot$, then we would expect of order 30
direct collisions within $b<2r_c$ over its lifetime, any of which would have
totally disrupted this cluster.

{\noindent (iv)} \ Finally, we note that $\tau_{_D}$ scales with galactocentric
distance as $r_{_G}^{2}$ due to the decreasing density of dark matter. Hence,
we expect the luminosity function of GCs to vary strongly with position as well
as with the mass of the parent galaxy.
However, the luminosity function of GCs does not appear to vary with position
either in our Galaxy or M31, neither is its shape observed to vary with the
mass
of the host galaxy (Harris 1991).

\medskip
	\medskip
\noindent{\bf Acknowledgments}\ \ I would like to thank Marc Davis, Kim Griest,
Ivan King and Joe Silk, and the referee of this letter for useful discussions
and comments. The Center for Particle Astrophysics at UC Berkeley and a
NATO/SERC fellowship are both acknowledged for their support.

	\beginref		%THIS WILL END DOUBLECOLUMN FORMAT

\pp Aguilar, L., Hut, P. and Ostriker, J.P. 1988, {\it Ap.J.}, {\bf 335}, 720.

\pp Barnes, J. and Hut, P. 1986, {\it Nature}, {\bf 324}, 446.

\pp Bennet, D. 1992, {\it Texas symposium}, Berkeley.

\pp Binney, J. and Tremaine, S. 1987, {\it Galactic Dynamics}, Princeton
University Press. Ed. J. Ostriker.

\pp Carr, B.J. 1979, {\it M.N.R.A.S.}, {\bf 189}, 123.

\pp Carr, B.J., Bond, J.R. and Arnett, W.D. 1984, {\it Ap.J.}, {\bf 277}, 445.

\pp Carr, B.J. and Lacey, C.G. 1987, {\it Ap.J.}, {\bf 316}, 23.

\pp Cen, R., Ostriker, J.P. and Peebles, P.J.E. 1993, {\it Princeton preprint.}

\pp Chandrasekhar, S. 1942, {\it Principles of Stellar Dynamics}, (Univ. of
Chicago press).

\pp Chernoff, D.F., Kochanek, C.S. and Shapiro, S.L. 1986, {\it Ap.J.},
{\bf 309}, 183.

\pp Chernoff, D.F. and Shapiro, S.L. 1987, {\it Ap.J.}, {\bf 322}, 113.

\pp Dicke, R.H. and Peebles, P.J.E. 1968, {\it Ap.J.}, {\bf 154}, 891.

\pp Djorgovski, S. 1986, {I.A.U. Symp.} {\bf 126}, 333.

\pp Djorgovski, S. and King, I.R. 1986, {\it Ap.J.Lett.}, {\bf 305}, L61.

\pp Fich, M. and Tremaine, S. 1992, {\it Ann.Rev.Astr.Astro.}, {\bf 29}, 409.

\pp Frenk, C.S., White, S.D.M., Efstathiou, G. and Davis, M. 1988, {\it Ap.J.},
{\bf 327}, 507.

%\pp Gould, A. 1992, {\it Ap.J.}, {\bf 392}, 442.

\pp Harris, W.E. 1991, {\it Ann.Rev.Astr.Astro.}, {\bf 29}, 543.

\pp Illingworth, G. and Illingworth, W. 1976, {\it Ap. J. Supp.}, {\bf 30},
227.

\pp Inman, R.T. and Carney, B.W. 1987, {\it Astro.J.}, {\bf 93}, 1167.

\pp King, I.R. 1966, {\it Astro.J.}, {\bf 71}, 64.

\pp Lacey, C.G. and Ostriker, J.P. 1985, {\it Ap.J.}, {\bf 299}, 633.

\pp Lee, H.M. and Ostriker, J.P. 1987, {\it Ap.J.}, {\bf 322}, 123.

\pp Magneville, C. 1992, {\it Texas symposium}, Berkeley.

\pp Palmer, P.L. and Papaloizou, J. 1987, {\it M.N.R.A.S.}, {\bf 224}, 1043.

\pp Peebles, P.J.E. 1987, {\it Nature}, {\it 327}, 210.

\pp Pryor, C., McClure, R.D., Fletcher, J.M. and Hesser, J.E. 1986,
{\it I.A.U. Symp.}, {\bf 126}, 661.

\pp Rees, M.J. 1987, {\it I.A.U. Symp.}, {\bf 117}, 395.

\pp Sandage, A. and Hartwick, F.D.A. 1977, {\it Astro.J.}, {\bf 82}, 459.

\pp Seitzer, P. and Carney, B.W. 1990, {\it Astro.J.}, {\bf 99}, 229.

\pp Spitzer, L. 1958, {\it Ap.J.}, {\bf 127}, 17.

\pp Trager, S.C., Djorgovski, S. and King, I.R. 1993, {\it Conf. proceedings:
The dynamics of Globular Clusters}, Berkeley.

\pp Webbink, R.F. 1984, {\it I.A.U. Symp.}, {\bf 113}, 541.

\pp White, R.E. and Shawl, S.J. 1987, {\it Ap.J.}, {\bf 317}, 246.

\pp Wielen, R. 1986, {\it I.A.U. Symp.}, {\bf 126}, 393.

\endref

\bye